\documentclass[printer]{aa}
\usepackage{graphics,epsfig,lscape,txfonts}
\usepackage{natbib}
\usepackage{german}
\bibpunct{(}{)}{;}{a}{}{,}

\def\cm-2{cm$^{-2}$}

\def\chandra{{\it Chandra}}

\def\xmm{{XMM-Newton}}
\def\m31{\object{M~31}}
\newcommand{\ergs}[1]{$\times 10^{#1}$ \hbox{erg s$^{-1}$}}
\newcommand{\oergs}[1]{$10^{#1}$ erg s$^{-1}$}
\newcommand{\hcm}[1]{$\times 10^{#1}$ cm$^{-2}$}

\newcommand{\expo}[1]{$\times 10^{#1}$}

\newcommand{\nh}{\hbox{$N_{\rm H}$}}

\begin{document}
\originalTeX

   \title{\xmm\ detection of type~I X-ray bursts in M~31\thanks{XMM-Newton is an ESA Science Mission 
    with instruments and contributions directly funded by ESA Member
    States and the USA (NASA).}
}

   \author{W.~Pietsch \and 
	   F.~Haberl
          }
\institute{Max-Planck-Institut f\"ur extraterrestrische Physik, 85741 Garching, Germany \\
%           e-mail: {\tt wnp@mpe.mpg.de}
           }
     
     \offprints{W.~Pietsch, \email{wnp@mpe.mpg.de}}
%     \mail{wnp@mpe.mpg.de}

   \date{Received 3 December 2004 / Accepted 15 December 2004}
   \titlerunning{Type~I X-ray bursts in M 31}

	\abstract{We searched for X-ray bursts in \xmm\ 
	archival data of X-ray sources in \m31\ globular clusters (GCs) and
	GC candidates. We detected two bursts simultaneously in EPIC pn and MOS
	detectors and some more candidates in EPIC pn. The energy distribution
	of the burst photons and the intrinsic luminosity 
	during the peak of the bursts indicate
	that at least the strongest burst was a type~I radius expansion burst. 
	The bursts identify the
	sources as neutron star low mass X-ray binaries in \m31. 
	The type~I X-ray bursts in \m31\ are the first 
	detected outside the Milky Way and show that with the help of \xmm\
	X-ray bursts can be used to classify neutron star low mass X-ray
	binaries in Local Group galaxies. 
	
} 
\maketitle

\section{Introduction}
Type~I X-ray bursts (hereafter X-ray bursts) were detected in low mass X-ray 
binaries (LMXBs) in the Milky Way and are explained as thermonuclear runaways in
freshly accreted material on the surface of a neutron star They may re-occur
on a time scale of hours to days, when enough new material is accreted,
compressed and heated, to densities and temperatures adequate for another
thermonuclear ignition
\citep[for reviews see][]{1993SSRv...62..223L,SB2003}. 
X-ray bursts show a large variety in profiles and peak flux. The persistent
luminosities of the sources when bursting, vary from $\sim$0.2 to well
bellow 0.01 of the Eddington luminosity.
Burst rise times last from less than a second to $\sim$10~s, and decay times are
in the range of $\sim$10~s to minutes. In general, burst profiles depend
strongly on photon energy, with decays which are much shorter at high photon
energies than at low energies. This energy dependence of the burst profile
corresponds to a softening of the burst spectrum during decay, which is the
result of cooling of the neutron star photosphere. The time dependent energy
spectra can be modeled by black body radiation of changing temperature.

During some bursts the energy
release is high enough that the luminosity at the surface of the neutron star
reaches the Eddington limit leading to an expansion of the neutron star
atmosphere (radius expansion X-ray bursts). These bursts show an energy
dependent
double peak structure. During expansion and subsequent contraction the
luminosity remains almost constant near the Eddington limit. However, the black
body temperature may quickly drop from more than 3 keV to well below 0.5 keV when the 
radius of the photosphere expands before it slowly rises again during contraction to
2--3~keV and then cools thereafter with continuously decreasing luminosity.
For a detailed description of the temperature and flux development of 
several radius expansion bursts from 
\object{4U/MXB 1820--30} see \citet{1987ApJ...314..266H}. 
The radius expansion bursts of about two thirds of the sources reach a
critical bolometric luminosity of $\sim$3.8\ergs{38}, corresponding to the 
Eddington limit for hydrogen-poor matter and may be used as standard
candles \citep[see e.g.][]{2003A&A...399..663K}. The expansion phase may last
for up to 30~s. 
To date thirteen luminous 
globular cluster (GC) X-ray sources are known in the Milky Way, twelve of which 
show X-ray bursts (and sometimes even radius expansion \hbox{X-ray} bursts) 
and therefore are believed 
to be LMXBs containing a neutron star \citep[see e.g.][]{VL2004}. The
distances of the sources vary between about 4 and 12 kpc.
Up to now, no X-ray bursts have been detected outside the Milky Way. 

With the high collecting power and good position resolution of the \xmm\ EPIC
cameras we now have the possibility to detect X-ray burst sources in all
galaxies of the Local Group. The X-ray burst signature then will identify 
these sources as neutron star LMXBs. Using the maximum burst flux as standard
candle would allow to determine the distance of the burst source. However, due 
to the limiting statistics, distance determinations by this method can not
compete with the accuracy of other methods.  The persistent
luminosity of many bursting LMXBs may be below the limiting point source 
sensitivity of current observations. 
Therefore, to find all X-ray burst sources in 
neighboring galaxies, one would have to search the entire extent of the galaxy 
with the resolution of the instrument point spread function (PSF).
A less complete method would be a search in LMXB candidates selected by other
means (e.g. as GC X-ray sources). Radius expansion bursts would be most easily 
picked up as they are the most luminous ``standard" type~I X-ray bursts. Besides
these bursts one may detect type~I ``superbursts" lasting for several hours 
with peak luminosities above \oergs{38} and maximum black body temperatures of
2--3 keV similar to the ones detected with BeppoSAX and RXTE from LMXBs in the
Milky Way \citep[see e.g.][]{SB2003}.

An ideal target for a search for X-ray burst sources is the bright Local Group
spiral \m31\ \citep[distance 780 kpc,][]{1998AJ....115.1916H,1998ApJ...503L.131S}
with its moderate Galactic foreground absorption  \citep[\nh = 7\hcm{20},
][]{1992ApJS...79...77S}. Many GC and GC candidates are known in \m31\ 
\citep[see e.g.][]{2004A&A...416..917G} which may host LMXBs. 
\citet[][hereafter PFH2005]{PFH2005} prepared a catalogue of \m31\
point-like X-ray sources analyzing all observations  in the \xmm\ archive which
overlap at least in part with the optical $D_{25}$ extent of the galaxy. 
In total, they detected 856 sources and identified and classified them using
their X-ray properties and correlations with sources in other wavelength
regimes. Twenty-seven of the X-ray sources are identified with GCs and ten with GC
candidates. They are -- extrapolating from bright GC sources in the Milky
Way -- most likely neutron star LMXBs.
In this paper we report on a search for X-ray bursts in the \m31\ \xmm\ data 
of these LMXB candidates.  
 
\section{XMM-Newton observations and analysis}
\begin{table}
\begin{center}
\caption[]{\xmm\ EPIC count rates for the pn and one MOS instrument
           (thin filter full frame mode, 15 arcsec on-axis extraction radius
	   corresponding to approximately the half energy width of the PSF) 
	   in different energy bands (in keV) expected from a source emitting a black 
	   body spectrum with temperature $T_{\rm bb}$ and intrinsic 
	   bolometric luminosity of \oergs{38} at the
	   \m31\ distance assuming Galactic foreground absorption of 7\hcm{20}.}
\begin{tabular}{rrrrrrr}
\hline\noalign{\smallskip}
\hline\noalign{\smallskip}
\multicolumn{1}{c}{$T_{\rm bb}$} & \multicolumn{3}{c}{EPIC pn} 
& \multicolumn{3}{c}{EPIC MOS} \\
& \multicolumn{3}{c}{0.2--0.5~~~0.5--4.5~~~4.5--7.0} & 
\multicolumn{3}{c}{0.2--0.5~~~0.5--4.5~~~4.5--7.0}  \\ 
\noalign{\smallskip}
(keV)&  \multicolumn{6}{c}{(10$^{-3}$ counts s$^{-1}$)} \\
\noalign{\smallskip}\hline\noalign{\smallskip}
0.1 & ~~~~124 & ~~~~125 &      0 &~~~~~~23 &      30 &      0 \\
0.2 &      70 &     364 &      0 &      14 &     109 &      0 \\
0.5 &      10 &	    333 &      2 &	 2 & ~~~~126 &	    1 \\
0.8 &       3 &     213 &     13 &       1 &      85 &      5 \\ 
1.0 &       2 &	    158 &     21 &	 0 &	  65 &	    8 \\
1.5 &       0 &      79 &     26 &       0 &      33 &      9 \\
2.0 &       0 &	     44 &     22 &	 0 &	  19 &	    8 \\
3.0 &       0 &	     17 &     12 &	 0 &	   7 &	    4 \\
\noalign{\smallskip}
\hline
\noalign{\smallskip}
\end{tabular}
\label{rate}
\end{center}
\end{table}

We searched for \m31\ burst sources in the same archival \xmm\ 
\citep{2001A&A...365L...1J}  EPIC 
\citep{2001A&A...365L..18S,2001A&A...365L..27T} observations that were used by
PFH2005
for the creation of the \m31\ source catalogue, i.e. pointings c1 to c4 to the
galaxy centre, n1 to n3 to the northern disk, s1 and s2 to the southern disk and
h4 to the northwest halo (see Table 1 of PFH2005 for details). The observations
were performed in the full frame mode (time resolution for pn and MOS 73.4~ms
and 2.6~s, respectively) using medium or thin filter with low background
exposure times of about 10 to 50 ks. For the light curve analysis we used
event pattern up to 12 (singles to quadruples).
We restricted our search to the stringent low 
background times used by PFH2005 to avoid spurious burst
candidates due to background flares which would show up in all EPIC instruments
at the same time and might be misidentified as bursts. 

To get an estimate for the counts that can be expected in the EPIC detectors
from an X-ray burst in \m31, we calculated count rates for typical burst spectra
(black body spectra with temperatures of 0.1 to 3 keV) for a source with an
intrinsic luminosity of \oergs{38} (Table~\ref{rate}). It is clear that only
bursts with radius expansion, that radiate at the Eddington luminosity  
for at least several seconds will give enough photons (more than about 10) 
to be significantly detected in pn detector and
possibly also several photons for burst confirmation in the MOS detectors. The 
expected number of photons in the individual MOS instruments is lower than for
the pn by a
factor of two or more depending on the assumed spectrum (see Table~\ref{rate}). 
The drop of
the black body temperature during the expansion to values below 1 keV
additionally favors their detection as it shifts the main energy output into the
band where EPIC is most sensitive. Taking all the points mentioned above into
account, a search strategy for bursts should be based on pn light curves of
candidate sources
with time resolutions of about 10~s or longer and use MOS count rates for
confirmation.

Therefore, we created light
curves for EPIC pn in the 0.5--4.5~keV band with 500~s, 100~s and 20~s
resolution for the 37 sources associated with a GC or possibly associated with a 
GC in the catalogue of PFH2005 selecting
all events within a radius around the source corresponding to the half energy
width of the PSF at the off-axis angle. We rejected light curves when the source was outside the EPIC pn 
field of view (FOV) or partly covered by EPIC pn CCD gaps. The effective
integrated exposure time for all sources adds up to 2.88\expo{6} s. 
We inspected the light curves by eye for burst candidates. Two bursts
are confirmed in MOS light curves and are therefore regarded as secure
detections. Background flares are ruled out as they would appear simultaneously
in other source light curves. The bursts most likely showed radius expansion and 
will be discussed in greater detail in the following sections. 
Several other pn burst candidates were not in the FOV of the MOS
detectors or not detected by them. 
No burst with characteristics 
similar to the superbursts in the Galaxy could be detected. However, as also in
the Galaxy superbursts are
very rare events, it is not surprising that none appears in the available
\m31\ observations.

The data analysis was performed using tools in the \xmm\ Science Analysis System
(SAS) v6.0.0, 
EXSAS/MIDAS 03OCT\_EXP, and 
FTOOLS v5.2 software packages, the imaging application DS9 v3.0b6 together with
the funtools package,
the mission count rate simulator WebPIMMS v3.6a
and the spectral 
analysis software XSPEC v11.3.1.   

\section{The X-ray burst source in the globular cluster [WSB85] S5 15}
\begin{figure}
   \resizebox{\hsize}{!}{\includegraphics[angle=-90]{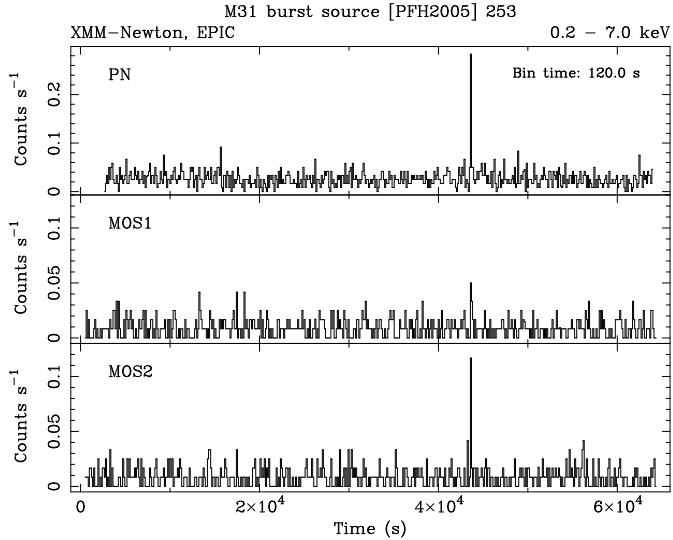}}
     \caption[]{
     \xmm\ EPIC light curves of source [PFH2005] 253 
     during the \m31\ centre observation c4 (OBSID 0112570101) on January 6/7, 
     2002 integrated over 120~s. Time zero corresponds to the start of the
     observation. 
     }
    \label{fig1} 
\end{figure}
\begin{figure}
   \resizebox{\hsize}{!}{\includegraphics[angle=-90]{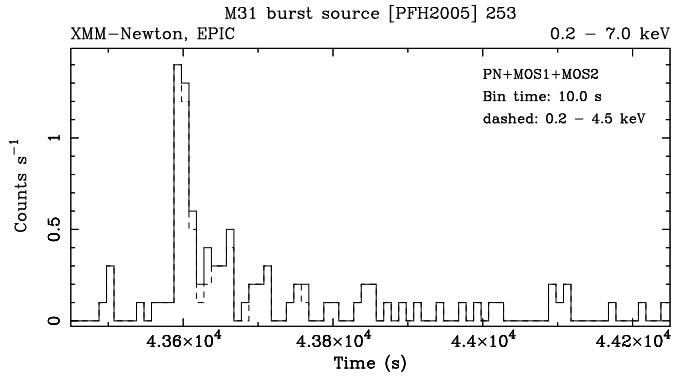}}
     \caption[]{
     Combined \xmm\ EPIC light curve (pn, MOS1 and MOS2 added) 
     of the burst in Fig~\ref{fig1} of 
     source [PFH2005]~253 integrated over 10~s. The dashed histogram shows the 
     light curve of the 0.2--4.5~keV events.  
     }
    \label{fig2} 
\end{figure}
\begin{figure}
   \resizebox{\hsize}{!}{\includegraphics[bb=94 30 565 700,angle=-90,clip]{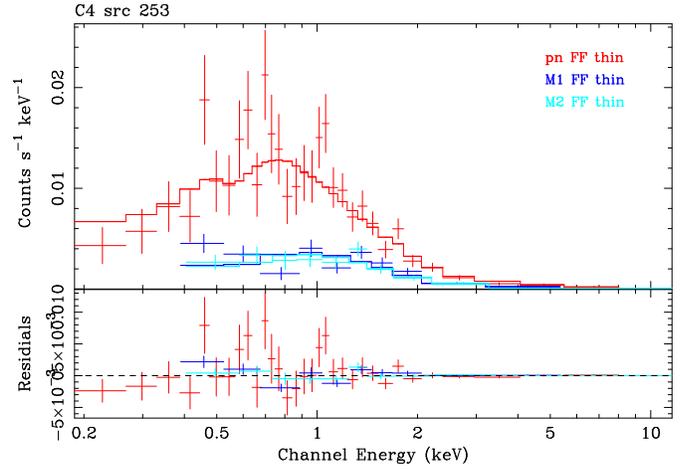}}
     \caption[]{
     \xmm\ EPIC spectrum of source [PFH2005] 253 for observation c4. 
     The best fit to the three EPIC spectra -- absorbed bremsstrahlung plus
     black body model -- is shown in the upper panel (pn is
     the upper spectrum and the two MOS spectra overlap). 
     }
    \label{fig3} 
\end{figure}

The brightest \m31\ burst in the EPIC pn and both MOS instruments 
was detected from source [PFH2005]~253 in GC [WSB85]~S5~15 
\citep{1985ApJ...290..140W} 
during observation c4 starting at UT 6:14:50 on January 7, 2002
(Fig.~\ref{fig1} and \ref{fig2}).
It reached its maximum luminosity within 10~s, stayed at the maximum 
(count rate in 0.2--7 keV band of (1.4$\pm$0.4)~ct~s$^{-1}$) for less than
20~s and decayed to background level within about 150~s. Due to limiting 
statistics the structure of the bursts can not be further resolved.
In total, more than 50 burst counts are detected. 
As can be seen from Fig.~\ref{fig2} photons in the 4.5--7 keV band are only
detected after the maximum of the burst as expected for the temperature
development in a radius expansion burst. Also the maximum count rate 
corresponds to that of a \m31\ source with a 1 keV black body spectrum radiating
at 3.8\ergs{38}, i.e. at the Eddington limit for hydrogen-poor matter.   

Source [PFH2005]~253 was in the EPIC FOV during the four \xmm\
\m31\ centre pointings c1 to c4 (average absorbed luminosity of 3.4\ergs{36} in 0.2--4.5
keV band). 
Between the pointings which are equally spread over two years, its average luminosity varied by a
factor of $\sim$6. During pointing c1 [PFH2005]~253 was partly hidden in pn by a CCD gap. We therefore only used
pointings c2, c3 and c4 for the burst search (effective low background time of 75.7 ks). 
 
During observation c4, more than 2\,500 photons were detected
in the EPIC detectors from [PFH2005]~253. An absorbed bremsstrahlung plus black 
body model describes the persistent energy spectrum (see Fig.~\ref{fig3}, best
fitting model parameters: 
\hbox{\nh = ($9^{+13}_{-5}$)\hcm{20},} \hbox{$T_{brems} = 5_{-2}^{+20}$ keV,} 
\hbox{$T_{bb} = 0.18_{-0.08}^{+0.06}$ keV,} reduced \hbox{$\chi^2_{\rm min}$ = 1.66} for 
45 degrees
of freedom) and corresponds to an absorbed luminosity and intrinsic luminosity
of the source in the 0.2--4.5 keV band of 4.1 and 5.6\ergs{36}, respectively.
The source was also detected by \chandra\ ACIS-I and HRC as a variable source
(r3-41, \citet{2002ApJ...577..738K}; CXOM31~J004221.5+411419, 
\citet{2002ApJ...578..114K}).

\section{The X-ray burst source in the globular cluster B150}
\begin{figure}
   \resizebox{\hsize}{!}{\includegraphics[angle=-90]{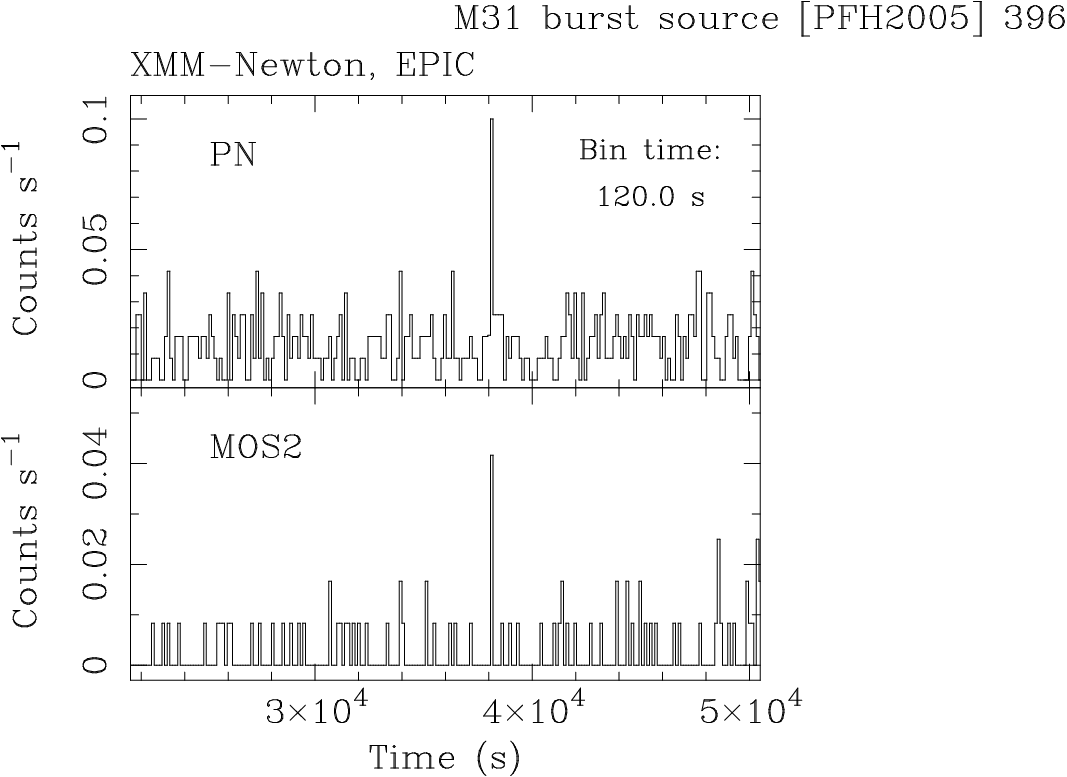}
                         \hspace{-4.7cm}
			 \includegraphics[angle=-90]{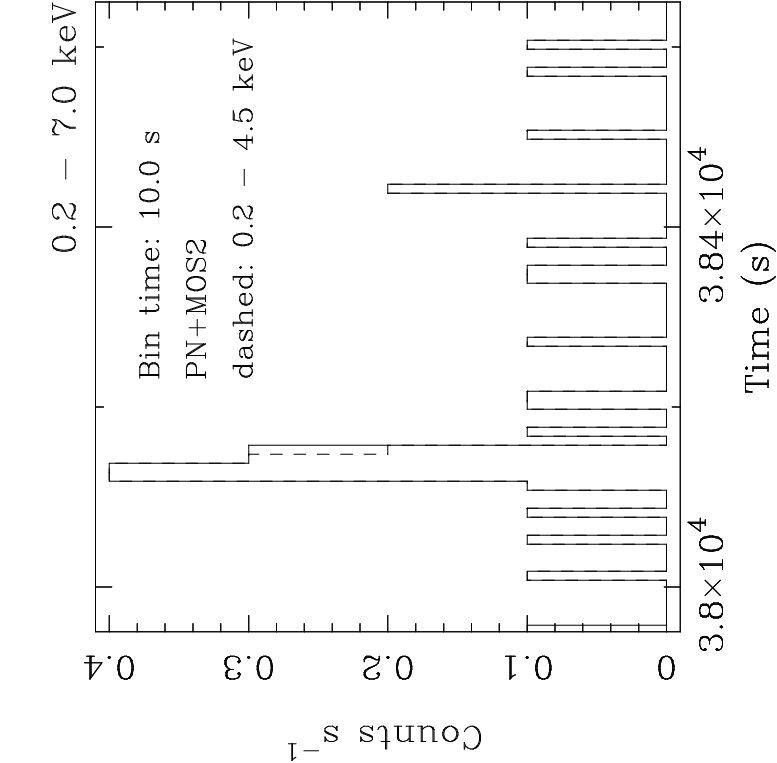}}
     \caption[]{Source [PFH2005] 396: 
     \xmm\ EPIC pn and MOS2 light curves 
     during the \m31\ centre observation c3 (OBSID 0112570701) on June 29, 
     2001 integrated over 120~s ({\bf left}). Combined \xmm\ EPIC light curve 
     (pn and MOS2 added) of the 
     burst integrated over 10~s ({\bf right}). Time zero corresponds to the
     start of the observation.      
     }
    \label{fig4} 
\end{figure}

The EPIC pn and the MOS2 instruments detected a burst from \m31\ source [PFH2005]~396 
in GC B150 \citep{2004A&A...416..917G} 
during observation c3 starting at UT 16:58:00 on June 29, 2001
(Fig.~\ref{fig4}). In MOS1 the source was heavily obscured by a CCD gap and
therefore MOS1 data could not be used for the analysis. The burst was
significantly fainter (about a factor of three) than the burst of [PFH2005]~253 and only about
15 counts are detected, four or them in MOS2. The burst lasted for about 40~s
with a rise to maximum within 10~s. Due to limiting 
statistics the structure of the bursts can not be further resolved. The count
rate at maximum provides us with a lower limit to the peak luminosity in \m31\ 
(assuming a 1 keV black body spectrum) of about \hbox{\oergs{38}}.

Source [PFH2005]~396 was in the EPIC FOV during the four \xmm\
\m31\ centre pointings c1 to c4 (average absorbed luminosity of 1.7\ergs{36} in 0.2--4.5
keV band). Between the pointings, its absorbed luminosity varied by a
factor of $\sim$1.7. We used pointings c1, c2, c3 and c4 for the burst search 
(effective low background time of 99.3 ks).

During observation c3, the source was at a luminosity of
(2.0$\pm$0.2)\ergs{36}. The faintness of the source and the low background
exposure time for observation c3 that was a factor of two shorter than for
observation c4, prevented a similar spectral
modeling as for [PFH2005]~253.  The source was also detected by \chandra\ ACIS-I and HRC as a variable source
(\hbox{r3-18,} \citet{2002ApJ...577..738K}; CXOM31~J004307.4+412021, 
\citet{2002ApJ...578..114K}). 

\section{Discussion}
We have detected the first type I X-ray bursts outside the Milky Way from 
two \m31\ GC source candidates of PFH2005 which showed persistent luminosities
while bursting of below $\sim$5\ergs{36}. In search for bursts we checked low 
background EPIC pn light curves (covering in total 2.88\expo{6} s) of
sources associated with GCs or possibly associated with GCs. The persistent 
luminosities of part of the sources was above 0.2 times the Eddington luminosity 
of a neutron star and therefore, extrapolating from the experience of Galactic 
LMXBs, were not expected to show bursts during these bright states. No burst with 
characteristics similar to the superbursts in the Galaxy was detected.

The peak luminosities of the two bursts in [WSB85]~S5~15 and B150 are
comparable to those of bright radius expansion bursts in the Milky Way. The
detections demonstrate the possibility to classify neutron star LMXBs in the
Local Group by their bursting signature using just X-ray information from 
\xmm\ EPIC. The energy response of the EPIC instruments is most sensitive to
detect radiation with black body temperatures in the range 0.2--1 keV which are
normally covered for some time during the contraction phase of the radius
expansion when the burst is radiating at the Eddington luminosity. 
Of specific importance is the possibility to simultaneously detect a burst
in more than one EPIC instrument, which gives high confidence in the credibility
of a burst even if only few photons are involved. 
In this work only known LMXB candidates in \m31\ from PFH2005 were investigated. 
The success encourages a more systematic survey  covering the total area of 
\m31\ and also other Local Group galaxies to identify neutron star LMXBs just from 
their bursting properties.
However, to be able to use the bursts as standard
candles as proposed by \citet[][]{2003A&A...399..663K} even better statistics 
at higher time resolution are needed to analyze the burst signature in more
detail.

\begin{acknowledgements}
The \xmm\ project is supported by the Bundesministerium f\"{u}r
Bildung und Forschung / Deutsches Zentrum f\"{u}r Luft- und Raumfahrt 
(BMBF/DLR), the Max-Planck Society and the Heidenhain-Stiftung.
\end{acknowledgements}

\bibliographystyle{aa}
\bibliography{./paper,/home/wnp/data1/papers/my1990,/home/wnp/data1/papers/my2000,/home/wnp/data1/papers/my2001}

\begin{thebibliography}{16}
\expandafter\ifx\csname natexlab\endcsname\relax\def\natexlab#1{#1}\fi

\bibitem[{{Galleti} {et~al.}(2004){Galleti}, {Federici}, {Bellazzini}, {Fusi
  Pecci}, \& {Macrina}}]{2004A&A...416..917G}
{Galleti}, S., {Federici}, L., {Bellazzini}, M., {Fusi Pecci}, F., \&
  {Macrina}, S. 2004, \aap, 416, 917

\bibitem[{{Haberl} {et~al.}(1987){Haberl}, {Stella}, {White}, {Gottwald}, \&
  {Priedhorsky}}]{1987ApJ...314..266H}
{Haberl}, F., {Stella}, L., {White}, N.~E., {Gottwald}, M., \& {Priedhorsky},
  W.~C. 1987, \apj, 314, 266

\bibitem[{{Holland}(1998)}]{1998AJ....115.1916H}
{Holland}, S. 1998, \aj, 115, 1916

\bibitem[{{Jansen} {et~al.}(2001){Jansen}, {Lumb}, {Altieri}, {Clavel}, {Ehle},
  {Erd}, {Gabriel}, {Guainazzi}, {Gondoin}, {Much}, {Munoz}, {Santos},
  {Schartel}, {Texier}, \& {Vacanti}}]{2001A&A...365L...1J}
{Jansen}, F., {Lumb}, D., {Altieri}, B., {et~al.} 2001, \aap, 365, L1

\bibitem[{{Kaaret}(2002)}]{2002ApJ...578..114K}
{Kaaret}, P. 2002, \apj, 578, 114

\bibitem[{{Kong} {et~al.}(2002){Kong}, {Garcia}, {Primini}, {Murray}, {Di
  Stefano}, \& {McClintock}}]{2002ApJ...577..738K}
{Kong}, A.~K.~H., {Garcia}, M.~R., {Primini}, F.~A., {et~al.} 2002, \apj, 577,
  738

\bibitem[{{Kuulkers} {et~al.}(2003){Kuulkers}, {den Hartog}, {in't Zand},
  {Verbunt}, {Harris}, \& {Cocchi}}]{2003A&A...399..663K}
{Kuulkers}, E., {den Hartog}, P.~R., {in't Zand}, J.~J.~M., {et~al.} 2003,
  \aap, 399, 663

\bibitem[{{Lewin} {et~al.}(1993){Lewin}, {van Paradijs}, \&
  {Taam}}]{1993SSRv...62..223L}
{Lewin}, W.~H.~G., {van Paradijs}, J., \& {Taam}, R.~E. 1993, Space Science
  Reviews, 62, 223

\bibitem[{{Pietsch} {et~al.}(2005){Pietsch}, {Freyberg}, \& {Haberl}}]{PFH2005}
{Pietsch}, W., {Freyberg}, M., \& {Haberl}, F. 2005, \aap, in press
  (astro-ph/0410117) (PFH2005)

\bibitem[{{Stanek} \& {Garnavich}(1998)}]{1998ApJ...503L.131S}
{Stanek}, K.~Z. \& {Garnavich}, P.~M. 1998, \apjl, 503, L131

\bibitem[{{Stark} {et~al.}(1992){Stark}, {Gammie}, {Wilson}, {Bally}, {Linke},
  {Heiles}, \& {Hurwitz}}]{1992ApJS...79...77S}
{Stark}, A.~A., {Gammie}, C.~F., {Wilson}, R.~W., {et~al.} 1992, \apjs, 79, 77

\bibitem[{{Str{\" u}der} {et~al.}(2001){Str{\" u}der}, {Briel}, {Dennerl},
  {Hartmann}, {Kendziorra}, {Meidinger}, {Pfeffermann}, {Reppin}, {Aschenbach},
  {Bornemann}, {Br{\" a}uninger}, {Burkert}, {Elender}, {Freyberg}, {Haberl},
  {Hartner}, {Heuschmann}, {Hippmann}, {Kastelic}, {Kemmer}, {Kettenring},
  {Kink}, {Krause}, {M{\" u}ller}, {Oppitz}, {Pietsch}, {Popp}, {Predehl},
  {Read}, {Stephan}, {St{\" o}tter}, {Tr{\" u}mper}, {Holl}, {Kemmer},
  {Soltau}, {St{\" o}tter}, {Weber}, {Weichert}, {von Zanthier},
  {Carathanassis}, {Lutz}, {Richter}, {Solc}, {B{\" o}ttcher}, {Kuster},
  {Staubert}, {Abbey}, {Holland}, {Turner}, {Balasini}, {Bignami}, {La
  Palombara}, {Villa}, {Buttler}, {Gianini}, {Lain{\' e}}, {Lumb}, \&
  {Dhez}}]{2001A&A...365L..18S}
{Str{\" u}der}, L., {Briel}, U., {Dennerl}, K., {et~al.} 2001, \aap, 365, L18

\bibitem[{Strohmayer \& {Bildsten}(2003)}]{SB2003}
Strohmayer, T. \& {Bildsten}, L. 2003, in Compact Stellar X-ray Sources, in
  press (astro--ph/0301544)

\bibitem[{{Turner} {et~al.}(2001){Turner}, {Abbey}, {Arnaud}, {Balasini},
  {Barbera}, {Belsole}, {Bennie}, {Bernard}, {Bignami}, {Boer}, {Briel},
  {Butler}, {Cara}, {Chabaud}, {Cole}, {Collura}, {Conte}, {Cros}, {Denby},
  {Dhez}, {Di Coco}, {Dowson}, {Ferrando}, {Ghizzardi}, {Gianotti}, {Goodall},
  {Gretton}, {Griffiths}, {Hainaut}, {Hochedez}, {Holland}, {Jourdain},
  {Kendziorra}, {Lagostina}, {Laine}, {La Palombara}, {Lortholary}, {Lumb},
  {Marty}, {Molendi}, {Pigot}, {Poindron}, {Pounds}, {Reeves}, {Reppin},
  {Rothenflug}, {Salvetat}, {Sauvageot}, {Schmitt}, {Sembay}, {Short},
  {Spragg}, {Stephen}, {Str{\"u}der}, {Tiengo}, {Trifoglio}, {Tr{\"u}mper},
  {Vercellone}, {Vigroux}, {Villa}, {Ward}, {Whitehead}, \&
  {Zonca}}]{2001A&A...365L..27T}
{Turner}, M. J.~L., {Abbey}, A., {Arnaud}, M., {et~al.} 2001, \aap, 365, L27

\bibitem[{Verbunt \& {Lewin}(2004)}]{VL2004}
Verbunt, F. \& {Lewin}, W.~H.~G. 2004, in Compact Stellar X-ray Sources, in
  press (astro--ph/0404136)

\bibitem[{{Wirth} {et~al.}(1985){Wirth}, {Smarr}, \&
  {Bruno}}]{1985ApJ...290..140W}
{Wirth}, A., {Smarr}, L.~L., \& {Bruno}, T.~L. 1985, \apj, 290, 140

\end{thebibliography}

\end{document}